\documentclass[numberedappendix]{emulateapj} 
\newcommand{\bl}[1]{\mbox{\boldmath$ #1 $}}

\slugcomment{Accepted by The Astrophysical Journal Letters}
\shorttitle{Crystalline silicate production}
\shortauthors{Vorobyov} 

\begin{document}

\title{Destruction of massive fragments in protostellar disks and crystalline silicate production}
\author{Eduard I. Vorobyov\altaffilmark{1,}\altaffilmark{2,}\altaffilmark{3}}
\altaffiltext{1}{Institute for Computational Astrophysics, Saint Mary's University,
Halifax, B3H 3C3, Canada; vorobyov@ap.smu.ca.} 
\altaffiltext{2}{Research Institute of Physics, Southern Federal University, Stachki 194, Rostov-on-Don, 
344090, Russia.} 
\altaffiltext{3}{Visiting Scientist, Department of Physics and Astronomy, The University of Western Ontario, London, 
ON N6A 3K7, Canada}


\begin{abstract}
We present a mechanism for the crystalline silicate production associated
with the formation and subsequent destruction of massive fragments in young protostellar disks. 
The fragments form in the embedded phase of star formation via disk fragmentation 
at radial distances $\ga 50-100$~AU and anneal 
small amorphous grains in their interior when the gas temperature exceeds the 
crystallization threshold of $\sim 800$~K. 
We demonstrate that fragments that form in the early embedded phase can be destroyed 
before they either form solid cores or 
vaporize dust grains, thus releasing the processed crystalline dust into various radial distances from sub-AU
to hundred-AU scales. 
Two possible mechanisms for the destruction of fragments are the tidal disruption and 
photoevaporation as fragments migrate radially inward and approach the central star 
and also dispersal by tidal torques exerted by spiral arms. 
As a result, most of the crystalline dust concentrates to the disk inner regions 
and spiral arms, which are the likely sites of fragment destruction.  
\end{abstract}


\section{Introduction}
The embedded phase of low-mass star formation, starting from the formation of a disk and ending with
the clearing of a parent cloud core, has recently drawn a renewed interest due to the possibility
of planet formation in this phase. Indeed, circumstellar disks are most massive in the embedded phase
\citep{Andrews05,Vor09a,Greaves10}
and numerical hydrodynamics simulations indicate that, under certain conditions, embedded disks
can fragment and form either protoplanetary embryos  or proto-brown dwarfs, 
especially in the disk outer regions where stellar irradiation and viscous heating are merciful
\citep[e.g.][]{Durisen07,Stamatellos09,Boley09,Machida10,VB10b,Cha10}.

While disk fragmentation in the embedded phase seems certainly feasible, the prospects for the 
survival of fragments and formation of gas giants or brown dwarfs on distances of order 
tens or hundreds AU are gloomy. 
Gravitational instability in the embedded phase is strong, fuelled with a continuing infall of gas from
a parent cloud core, and resultant gravitational and tidal torques are rampant. 
As a consequence, part of the fragments are dispersed by the tidal torques exerted by 
the spiral arms \citep{VB10b,Boley10}.
Others are quickly driven into the disk inner regions, and probably
onto the star, due to the loss of angular momentum caused by the gravitational interaction 
with the trailing spiral arms \citep{VB06,VB10b,Cha10}. 
Only those fragments that happen to form in the late embedded phase, when 
gravitational instability and associated torques are getting weaker,
may open a gap in the disk and  mature into gas giants on wide orbits but the probability for such 
a ''fortunate'' outcome is rather low \citep{VB10b}.

Ironic as it may seem, but the destruction of fragments (hereafter, planet/brown-dwarf embryos or simply
embryos) may be as important for the disk and 
stellar evolution as their survival followed by massive giant planet or brown dwarf formation.
Embryos that are driven into the disk inner regions may form solid cores in their interior via 
dust sedimentation or dense and compact atomic hydrogen cores via dissociation 
of molecular hydrogen. If one or both of those processes take place before the embryo
is destroyed on its approach to the central star, 
a planet can emerge on orbit of order several AU or less \citep{Nayakshin10,Boley10}. 
If both processes 
fail, then the release of gravitational energy from the accreted embryo can trigger a luminosity outburst
similar in magnitude those of FU Orionis or EX~Lupi family \citep{VB06,VB10b}. 
In addition, short but intense bursts of accretion onto the star caused by the consumption of embryos
can lower the lithium abundance in young
solar-type stars \citep{Baraffe10}, explain the luminosity spread in H-R diagrams of 1-Myr-old
star clusters \citep{Baraffe09} and resolve the long-standing luminosity
problem \citep{Dunham10}. 

On the other hand, embryos are sites of accelerated dust growth and high-temperature
processing \citep{Boss98,Boss02,Boley10,Nayakshin10, Nayakshin10b} and
their in situ destruction by tidal torques may release processed dust directly into the disk 
outer regions.

Whichever mechanism of embryo destruction is realized in nature, the consequences of this process
are far-stretching.
In this Letter, we report another interesting by-product of the embryo dispersal---the enrichment
of embedded circumstellar disks with crystalline silicates formed in the depths of 
massive embryos via thermal annealing of amorphous dust grains. We use numerical hydrodynamics 
simulations of the disk formation
and long-term evolution complemented with equations describing the transformation of 
micron-sized amorphous dust grains into the silicate form.  We show that part of 
the crystalline-silicate-bearing embryos are destroyed before they vaporize dust grains in their interiors
or form solid cores via dust sedimentation, thus enriching the gas disk with crystalline silicates 
on radial distances from sub-AU to hundred-AU scales.

\section{Sources of crystalline silicates in circumstellar disks}
Circumstellar disks originate during the gravitational collapse of rotating prestellar 
cloud cores.
The degree in crystallinity of silicates in prestellar cores is similar to that 
of the interstellar medium,
which has an upper limit of $0.2\pm 0.2\%$ in the direction of the Galactic center
\citep{Kember04}. On the other hand, the crystalline-silicate fraction in the inner regions of
circumstellar disks 
around young stellar objects (YSOs) can vary in wide limits from essentially none
to almost 100\% \citep{Watson09}, implying in situ production of crystalline silicates in at
least some YSOs. Two mechanisms for dust crystallization have been put forward. 
\begin{enumerate} 
\item Evaporation of the original, 
amorphous dust grains followed by re-condensation under conditions of high temperature  
and density \citep[e.g.][]{Grossman72}
\item
Thermal annealing of the amorphous grains at temperatures (800--1300)~K, somewhat below 
their vaporization point, via viscous heating \citep[e.g.][]{Gail01},
shock wave heating \citep{Harker02}, 
or disk surface heating during EX-Lupi-like outbursts \citep{Abraham09}. 
\end{enumerate}
Both mechanisms are thought to work mostly in the inner 1~AU from 
the star (except for shock heating that may operate at somewhat larger radial
distances $r\la 10$~AU) and some 
means of outward radial transport of crystallized dust is necessary to account for 
the non-zero crystalline-silicate fraction in comets at distances of order tens of AU 
\citep[e.g.][]{Gail01,Bockelee02,Ciesla07}.

\section{Model description and initial conditions} 
Our numerical model is explained in detail in \citet{VB10b}. Here, we provide only a brief
overview and describe modifications done to the numerical code. We compute the gravitational 
collapse of rotating cloud cores in the thin-disk approximation 
and start our simulations from the pre-stellar phase, advance 
through the disk formation and early evolution phase, and terminate 
with almost a complete clearing of the natal core. 
The following physical effects are taken into account in our model: 
stellar irradiation, background irradiation with temperature $T_{\rm bg}=10$~K, 
viscous and shock heating, radiative cooling from the
disk surface, and disk self-gravity. Turbulent viscosity is parameterized using the usual 
$\alpha$-prescription. A spatially and temporally constant $\alpha$ is set to 0.005.

In the current version, we have implemented a simple mechanism describing the transformation 
of amorphous dust grains into the crystalline form following the guidelines described
in \citet{Dullemond06}.
We assume that dust is passively transported with gas and neglect the dust 
radial drift caused by the dust-gas drag force. The latter simplification 
is of little consequence for the dynamics of dust grains with sizes $\la 10~\mu m$  on timescales
of interest for the present paper, $\la 0.5$~Myr \citep{Takeuchi02}.
Therefore, our results are strictly applicable only to sufficiently small dust grains showing
sharp features in the 5--35~$\mu m$ band of the mid-infrared spectrum.
In this simple scheme, two additional continuity equations for the surface density of amorphous 
and crystalline silicates, $\Sigma_{\rm a.s.}$ and $\Sigma_{\rm c.s.}$, respectively, are 
\begin{eqnarray}
\label{amsil}
{\partial \Sigma_{\rm a.s.} \over \partial t} &+& \nabla_p  \cdot 
\left( \Sigma_{\rm a.s.} \bl{v}_p \right) = -S, \\
\label{crsil}
{\partial \Sigma_{\rm c.s.} \over \partial t} &+& \nabla_p  \cdot 
\left( \Sigma_{\rm c.s.} \bl{v}_p \right) = +S,
\end{eqnarray}
where $\bl{v}_{p}=v_r \hat{\bl r}+ v_\phi \hat{\bl \phi}$ is the gas velocity in the
disk plane and $S=\nu_{\rm cr} \Sigma_{\rm a.s.}$. 
All dust grains in a collapsing cloud core are initially amorphous, they
crystallize at a rate $\nu_{\rm cr}$ provided by the Arrhenius formula
\begin{equation}
\nu_{\rm cr}\left[s^{-1}\right]=2\times 10^{13} \exp\left({-{T_{\rm c}\over T_{\rm g}}}\right),
\label{Arrhenius}
\end{equation}
where $T_{\rm g}$ is the gas midplane temperature, $2\times 10^{13}$ is the lattice vibrational
frequency, and $T_{\rm c}$=38100~K \citep{Bockelee02}. We assume that dust and gas are 
thermalized. The initial gas and dust radial surface density profiles and angular velocities 
are given by equations~(12) and (13) in \citet{VB10b} and
the initial dust-to-gas ratio is 0.01. The initial core mass is $M_{\rm core}=0.78~M_\odot$ and 
the ratio of rotational to gravitational energy of the core is $\beta=8.8\times 10^{-3}$. 
For simplicity, we also assume that all dust is made of silicate grains. 

A solution of the time-dependent part of Equation~\ref{crsil} (excluding advection terms) 
demonstrates that it takes about 40~yr to completely transform amorphous dust
into the crystalline form at $T_{\rm g}=800$~K. The corresponding timescale for $T_{\rm g}=900$~K
is 0.1~yr, while for $T_{\rm g}=$700~K the crystallization timescale is 20000~yr. 
Obviously, the latter value is much longer than the typical orbital period and 
crystallization temperatures of $T_{\rm cr}=800-900$~K are most relevant to circumstellar disks.
Laboratory experiments tend to yield even higher $T_{\rm cr}\ga 1000$~K \citep{Hallenbeck00}, possibly due to the
same timescale arguments.

If temperature exceeds the evaporation threshold for dust grains 
$T_{\rm evap}\approx 1400$~K, we set both $\Sigma_{\rm c.s.}$ and $\Sigma_{\rm a.s.}$ 
to zero in the corresponding computational cell. The exact value of $T_{\rm evap}$ is density
dependent and is calculated from the opacity tables of \citet{Bell94}.

Equations~(\ref{amsil}) and (\ref{crsil}) are solved simultaneously with 
equations of hydrodynamics describing the formation and evolution of circumstellar disks
\citep[eqs (1)--(3) in][]{VB10b}. We employ the method of finite 
differences with a time-explicit, operator-split solution procedure in polar coordinates 
$(r, \phi)$ on a numerical grid with $512 \times 512$ grid zones.
The advection part of Equations~(\ref{amsil}) 
and (\ref{crsil}) (i.e., with $S$ set to zero) is solved using a piece-wise 
parabolic advection scheme. The time update of $\Sigma_{\rm a.s.}$  and $\Sigma_{\rm c.s.}$ 
due to the source/sink term $S$ is done using an (unconditionally stable) backward difference 
scheme supplemented with a subcycling procedure to guarantee accuracy, wherein the change 
in both $\Sigma_{\rm a.s.}$ and $\Sigma_{\rm c.s.}$ over one global hydrodynamical timestep  
is limited to 10\%. If this condition is violated, local subcycling with smaller timesteps 
is invoked until the desired accuracy is reached.

\begin{figure*}
 \centering
  \includegraphics[width=18cm]{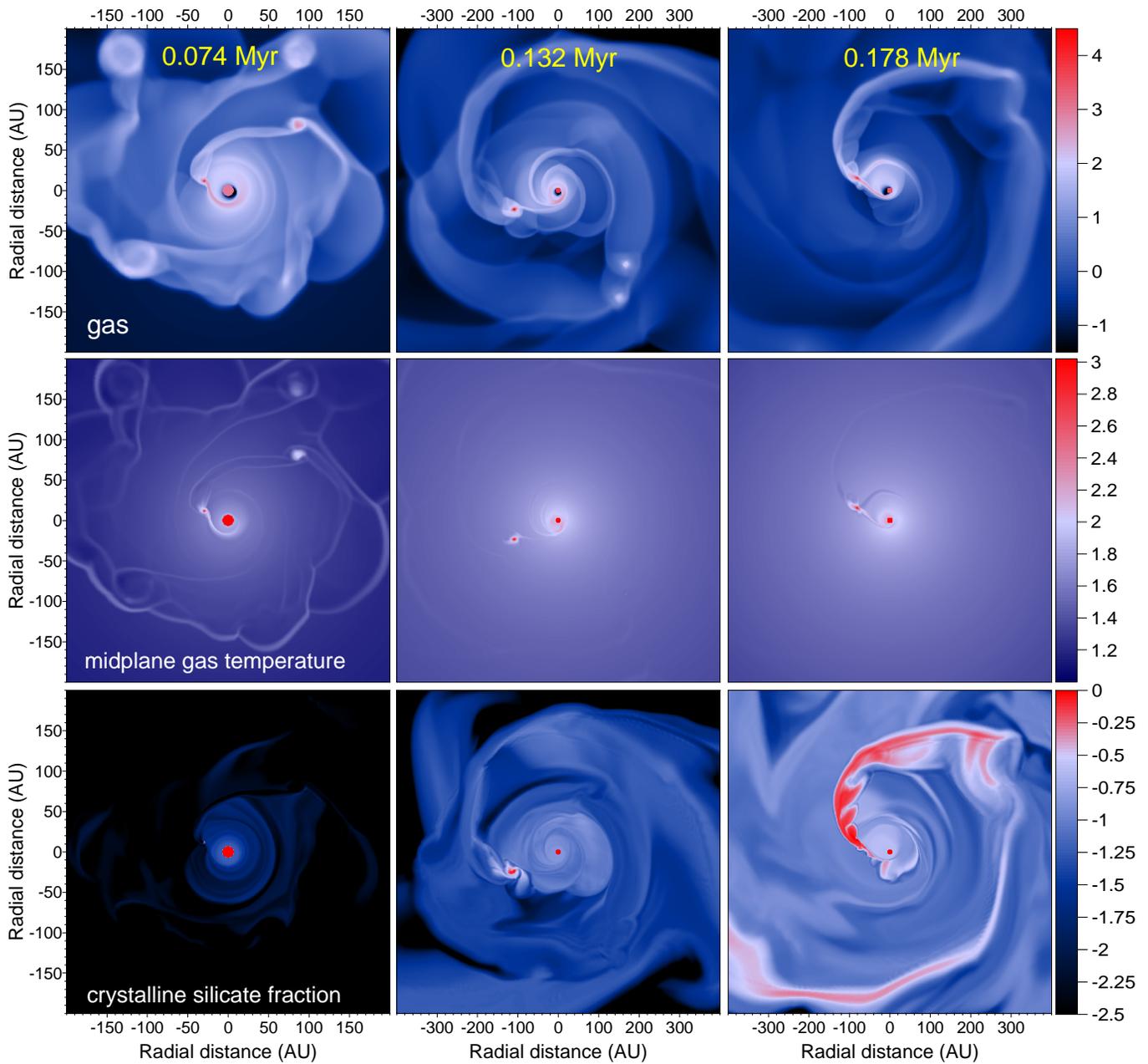}
      \caption{Gas surface density (g~cm$^{-2}$, top row), gas midplane temperature 
      (Kelvin, middle row),
      and crystalline silicate fraction (bottom row) at three times since the formation of the central
      star: $t=0.074$~Myr
      (left column), $t=0.132$~Myr (middle column), and $t=0.178$~Myr (right column).}
         \label{fig1}
\end{figure*}

The radial points in our numerical grid are logarithmically spaced.
The innermost grid point is located at the position of the sink cell $r_{\rm sc}=6$~AU and the 
size of the first adjacent cell is 0.08~AU. This corresponds to the radial 
resolution of $\triangle r$=1.3~AU at 100~AU.
We impose a free outflow boundary condition so that
the matter is allowed to freely flow through the sink cell to form a central star plus some 
dynamically inactive inner region.

\section{Results}

The top row in Figure~\ref{fig1} presents the gas surface density in the inner several
hundred AU at three distinct evolution times since the formation of the central star.
The middle and bottom rows show the corresponding gas midplane temperature and crystalline
silicate fraction $\xi_{\rm c.s.}=M_{\rm c.s.}/(M_{\rm c.s.}+M_{\rm a.s.})$, respectively,
where $M_{\rm c.s}$ and $M_{\rm a.s.}$ are crystalline and amorphous silicate masses.
The red circle in the coordinate center represents schematically the central star plus sink cell.
The three time snapshots in Figure~\ref{fig1} are chosen specifically to illustrate
disk fragmentation and embryo formation (red dots at 50--100~AU). We note that
each snapshot shows in fact different embryos---their inward migration and/or tidal destruction
timescales are of order several orbital periods ($T_{\rm orb}=700$~yr 
for $M_\ast=0.25~M_\odot$, $r=50$~AU, $t=0.074$~Myr and $T_{\rm orb}=1600$~yr
for $M_\ast=0.4$, $r=100$~AU, $t=0.178$~Myr), 
which are considerably shorter than the time span between the images ($\sim 5\times 10^4$~yr). 

We find that the gas temperature $T_{\rm g}$ in the interiors of 
massive embryos can exceed $800$~K (see middle row in Figure~\ref{fig1}), 
necessary for the thermal annealing of amorphous dust grains to commence. 
The prospects for disk enrichment with crystalline silicates stored in these embryos
will depend on characteristic timescales for inward migration and/or tidal destruction of embryos
($\tau_{\rm dest}$), dust sedimentation and solid core formation in the embryo interiors 
($\tau_{\rm sed}$), and dust vaporization at $T_{\rm g}>T_{\rm evap}$ ($\tau_{\rm evap}$). 
If $\tau_{\rm dest}$ is smaller than both $\tau_{\rm sed}$ and $\tau_{\rm evap}$,
then all processed dust will be released into the disk at various radial distances, depending on which
destruction mechanism prevails. As current numerical simulations indicate, embryos may be 
destroyed by tidal torques from spiral arms, preferentially releasing the crystalline silicates to large
radial distances where these embryos form. Alternatively, embryos may be photoevaporated 
and torn apart by stellar irradiation and tidal torques as they migrate radially inward and 
approach the central star. 
In this case, part of the processed dust will be brought into the disk inner few AU (through the sink cell) and probably onto the star and part 
will be pushed to larger disk radii during a transient episode of disk expansion (caused by the conservation
of angular momentum) following the embryo tidal disruption. It
is thus feasible to deposit crystalline silicates to various orbital distances starting from sub-AU
scales and up to several hundred
AU, accounting for the crystalline silicate features observed in the spectra of the Solar system comets.
This process of crystalline silicate enrichment is illustrated in the bottom row of 
Figure~\ref{fig1}---there is a notable increase in $\xi_{\rm c.s.}$ with time. 
Crystalline silicates concentrate in the disk inner regions and at/near spiral arms, 
which are likely sites of embryo formation and destruction.
This implies that the abundance of crystalline silicates may increase with radius, at least in
the early disk evolution, if most of the embryos are tidally destroyed in the disk outer regions rather
than driven into the star. 
If such an increase is observationally confirmed, it could become a hallmark for the proposed process. 

On the other hand, if $\tau_{\rm dest}$ greater than either $\tau_{\rm sed}$ or $\tau_{\rm evap}$, then
most of the processed crystalline dust will be either locked up in solid terrestrial-like cores or vaporized.
The latter mechanism is implemented in the current numerical model, while the effect of the former
is discussed in more detail later in the text.

\begin{figure*}
 \centering
  \includegraphics[width=16cm]{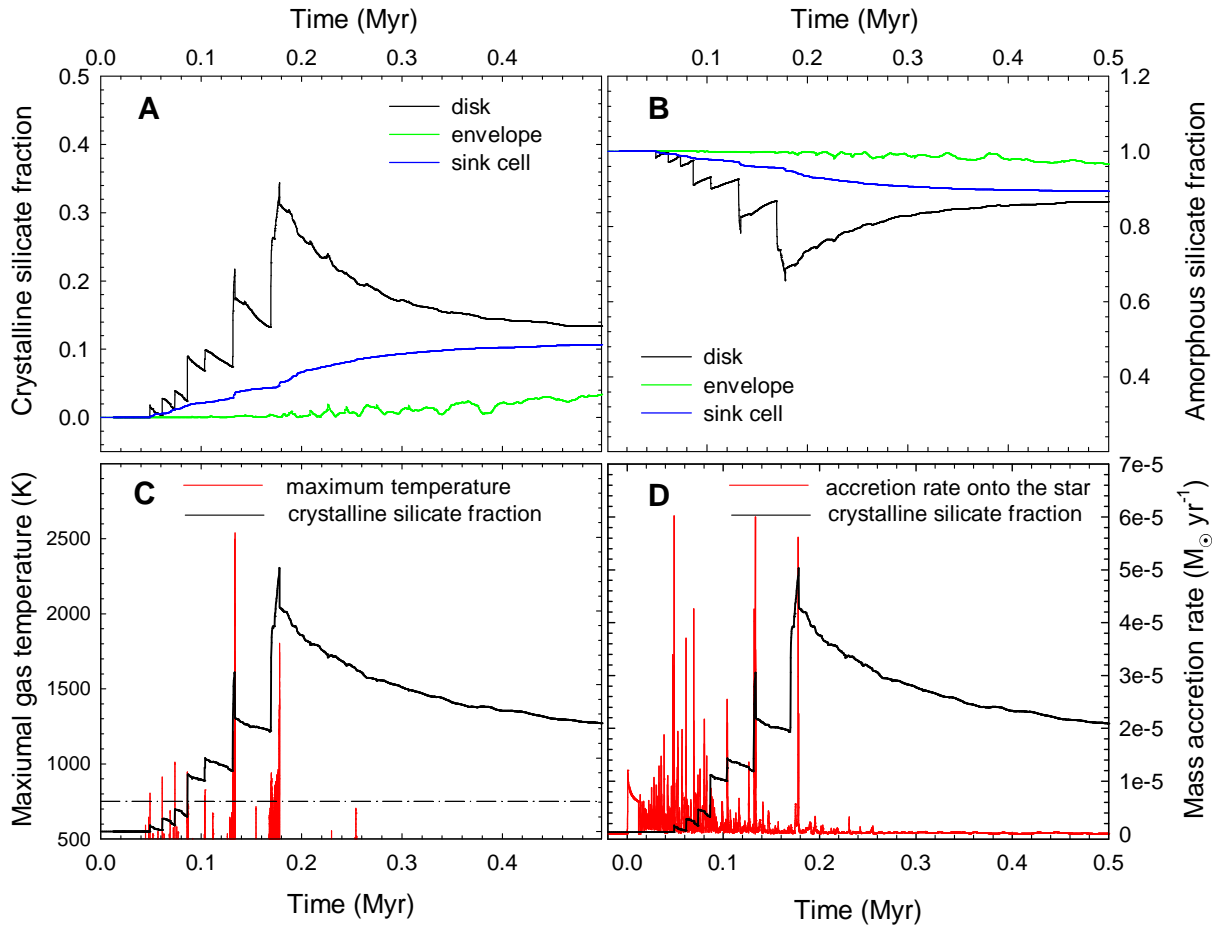}
      \caption{Crystalline (panel~A) and amorphous (panel~B) silicate fractions in 
      the disk (black), envelope (green), and star+sink cell (blue) as a function of time passed since
      the formation of the central star. Panel~C shows the maximum temperature in the disk (red) and crystalline
      silicate fraction in the disk (black), while panel~D also shows the mass accretion rate
      onto the star (red). }
         \label{fig2}
\end{figure*}

To better illustrate the process of disk enrichment with crystalline silicates,
Figure~\ref{fig2} presents various model characteristics as a function of time. 
In particular, the crystalline silicate fraction 
$\xi_{\rm c.s.}$ and amorphous silicate fraction 
$\xi_{\rm a.s.}=M_{\rm a.s.}/(M_{\rm c.s.}+M_{\rm a.s.})$ are
plotted in panels~A and B, respectively.  More specifically,
black, green, and blue lines depict the corresponding fractions in
the disk ($\xi_{\rm c.s.}^{\rm disk}$ and $\xi_{a.s.}^{\rm disk}$), 
envelope ($\xi_{\rm c.s.}^{\rm env}$ and $\xi_{a.s.}^{\rm env}$), and sink cell
($\xi_{\rm c.s.}^{\rm sc}$ and $\xi_{a.s.}^{\rm sc}$), respectively. 
We distinguish between the disk and envelope using a scheme based on the typical
transitional gas surface density  and radial gas velocity field
explained in detail in \citet{Vor10}. We note that $\xi_{\rm c.s.}^{\rm disk}$ and $\xi_{\rm a.s.}^{\rm
disk}$ are in fact crystalline and amorphous silicate fractions stored in both the disk and the embryos.
Panel~C shows the crystalline silicate fraction in the disk (black line) and maximum 
midplane gas temperature $T^{\rm max}_{\rm g}$ (red line), whereas panel~D presents
the mass accretion rate onto the star $\dot{M}$ (red line) and the crystalline silicate 
fraction in the disk (black line).

There are seven episodes of crystalline silicate production in the disk manifested by a sharp increase
in $\xi_{\rm c.s.}^{\rm disk}$, all associated with
the formation of embryos massive and hot enough to crystallize dust in their interior. 
Episodes 4, 6, and 7 (from left to 
right) correspond to the formation of embryos shown in the left, middle, and right columns 
of Figure~\ref{fig1}. From panel~C
it is evident that $\xi_{\rm c.s.}^{\rm disk}$ increases sharply each time the maximum gas temperature
$T_{\rm g}^{\rm max}$ exceeds the crystallization temperature $T_{\rm cr}\approx 800$~K 
(horizontal dash-dotted line). After each such episode, prolonged periods of gradual decline 
in $\xi_{\rm c.s.}^{\rm disk}$ 
follow, which are caused by dilution of the disk material with pristine amorphous dust 
accreted onto the disk from the infalling envelope.

In the last two episodes, $T_{\rm g}^{\rm max}$ goes above the dust evaporation 
temperature $T_{\rm evap}$  and
this leads to an immediate sharp drop in $\xi_{\rm c.s.}^{\rm disk}$. 
This exemplifies cases with $\tau_{\rm evap}<\tau_{\rm dest}$. 
The dust evaporation 
timescale can be approximated (excluding factors of unity) as \citep{Nayakshin10}
\begin{equation}
\tau_{\rm evap}=1.5\times 10^4 \left( {M_{\rm emb} \over 10 \, M_{\rm J} }\right)^{-2} \,\, {\rm yr}.
\label{tevap}
\end{equation}
For the typical embryo masses of $M_{\rm emb}$=5--20~$M_{\rm J}$, 
the corresponding $\tau_{\rm evap}$
lies in the 4000--60000~yr range. In fact, this is a lower estimate 
simply because embryos may experience several contraction 
and expansion episodes before actually reaching $T_{\rm evap}$.
Embryos that form in the early disk evolution are of lower mass and closer to the star than
those that form later in the embedded phase (because disk grows in mass and size with time).
As a consequence, first embryos take longer time to evaporate dust and shorter time to migrate 
into the star, an effect found in our numerical simulations and also noted by \citet{Cha10}.
For these first embryos, migration timescales are just several orbital periods 
($T_{\rm orb}$=700--1600~yr) 
and are often shorter than $\tau_{\rm evap}$. As time passes and more massive embryos 
on wider orbits start to form, 
the migration/destruction time scale may exceed that of dust evaporation, as indeed occurred
during the last two episodes of crystalline dust production.

Migration of embryos into the disk inner regions and through the sink cell, 
as manifested by strong 
($\ga 5\times 10^{-5}~M_\odot$~yr$^{-1}$) mass accretion bursts, also contributes to 
the decrease in $\xi_{\rm c.s.}^{\rm disk}$ but leads to an increase in $\xi_{\rm c.s.}^{\rm sc}$.
Note that episodes 3 and 4 are not associated with strong mass accretion bursts,
indicating that the corresponding embryos are dispersed by tidal torques exerted by spiral arms,
before they actually fall through the sink cell
onto the star. The process of disk fragmentation diminishes after $t\approx0.2$~Myr, no embryos survive
beyond this time, and
the subsequent evolution shows a gradual decline in $\xi_{\rm c.s.}^{\rm disk}$. From this moment on,
$\xi_{\rm c.s.}^{\rm disk}$ represents the crystalline silicate fraction in the disk only. 

We do not take into account a possible formation of solid cores (and associated reduction in 
disk enrichment with crystalline dust) but we can estimate its efficiency 
based on the work of \citet{Nayakshin10}. For $M_{\rm emb}=3-10~M_{\rm J}$, Nayakshin found
$\tau_{sed}\approx (4-6)\times 10^3$~yr. This is comparable to or less than 
$\tau_{\rm evap}\ga(4-16)\times10^3$~yr, which implies that solid cores may form before dust is 
vaporized. At the same time, 
$\tau_{\rm sed}$ is comparable to or longer than $\tau_{\rm dest}=\mathrm{a~few} \times T_{\rm orb}\approx
\mathrm{a~few} \times 10^3$~yr, 
certainly for those embryos that form first in the early embedded phase and are characterized 
by shortest migration timescales.
This simple timescale analysis suggests that the solid core formation is feasible and may 
reduce the efficiency of crystalline dust enrichment but is unlikely to shut down the proposed 
mechanism completely. The latest numerical hydrodynamics simulations
seem to confirm that the solid core formation is not expected to work for every
embryo in the disk \citep{Cha10}.

\section{Summary}
Thermal annealing of amorphous dust grains in the hot interiors of massive fragments (or planet/brown
dwarf embryos) forming in protostellar disks during the early embedded stages of stellar 
evolution, followed by subsequent dispersal of the fragments,
is a promising gateway for the production of crystalline silicates. This mechanism can release processed
dust to various radial distances from (sub-)AU scales, if fragments are destroyed on their approach
to the central star, to hundred-AU scales, if they are dispersed by tidal torques exerted by spiral
arms. More studies, including the effect of dust growth and solid core formation, 
are needed to further explore the efficiency of this mechanism.

We have run another eight models with $\beta=(3.2-12)\times10^{-3}$ and 
$M_{\rm core}=0.3-1.25~M_\odot$. The net result is that the dust processing is correlated with disk
fragmentation and models that do not form embryos (e.g., models with low $M_{\rm core}$ 
and $\beta$ and consequently with disks of low mass and size)
show little crystalline silicates, reinforcing our conclusions. 

We stress that our proposed mechanism can coexist with other 
crystallization mechanisms, localized to the inner few AU, and complements them by providing 
a direct source of crystalline silicates at large distances.  
Some outward transport may still takes place and account for the fact that comet Wild~2 has  
crystalline silicates with the chemical and isotopic composition similar to that of the 
chondritic meteorites originated in the terrestrial
planet region  \citep{McKeegan06}.
Conversely, chondritic meteorites may 
have formed in the atmospheres of protoplanetary embryos via dust sedimentation  
and released into the inner few AU when the embryos were dispersed by tidal torques near the young Protosun.
The apparent presence of refractory Calcium Aluminum Inclusions in the cometary material does not
invalidate this scenario since temperatures in at least some of the embryos can exceed 1300~K 
(panel~C in Figure~\ref{fig2}). We do not take into account possible opacity dependence 
on the composition and properties of dust grains due to the complexity of physics involved. 
Nevertheless, dust growth and crystallization are likely to 
decrease opacities \citep{Mennella98,DAlessio01}, thus enhancing disk fragmentation and 
crystalline silicate production.

\acknowledgements
The author is grateful to the anonymous referee for very useful comments and 
to Prof. Shantanu Basu for hospitality. Support from an ACEnet Fellowship
is gratefully acknowledged. Numerical simulations were done 
on the Atlantic Computational Excellence Network (ACEnet).
This project was also supported by RFBR grant 10-02-00278 and by the 
Ministry of Education grant RNP 2.1.1/1937.

\end{document}